\documentclass{article}

\pdfoutput=1

\usepackage{twocolceurws}
\usepackage{times}
\usepackage{color}
\usepackage{graphicx}

\title{A Scalable Data Streaming Infrastructure for Smart Cities}

\author{
Jes\'us Arias Fisteus\textsuperscript{\textnormal{1}},
Luis S\'anchez Fern\'andez\textsuperscript{\textnormal{1}},
V\'ictor C\'orcoba Maga\~na\textsuperscript{\textnormal{1}},
Mario Mu\~noz Organero\textsuperscript{\textnormal{1}}, \\
Jorge Yago Fern\'andez\textsuperscript{\textnormal{2}},
Juan Antonio \'Alvarez-Garc\'ia\textsuperscript{\textnormal{2}} \\
\textsuperscript{1}Depto. de Ingenier\'ia Telem\'atica, Universidad Carlos III de Madrid \\
\{jaf, luiss, vcorcoba, munozm\}@it.uc3m.es \\
\textsuperscript{2}Depto. de Lenguajes y Sistemas Inform\'aticos, Universidad de Sevilla \\
\{jorgeyago, jaalvarez \}@us.es
}

\institution{}

\begin{document}

\maketitle

\begin{abstract}
Many of the services a smart city can provide to its citizens
rely on the ability of its infrastructure
to collect and process in real time
vast amounts of continuous data
that sensors deployed through the city produce.
In this paper we present the server infrastructure
we have designed in the context of the HERMES project
to collect the data from sensors
and aggregate it in streams
for their use in services of the smart city.
\end{abstract}

\section{Introduction}

Many of the services a smart city can provide to its citizens
rely on the ability of its infrastructure
to collect and process in real time
vast amounts of continuous data
that sensors deployed through the city produce~\cite{perera2014}.
In this scenario,
building an infrastructure that scales
as the number of such sensors and their data rates increase
is a challenging task.
Grouping the data in streams is a common approach
for this kind of scenarios.
A data stream can be defined as \emph{a real-time,
  continuous, ordered (implicitly by arrival time or explicitly by
  timestamp) sequence of items}~\cite{Golab2003}.
Streams are different to stored data
in several aspects: they cannot normally be stored in their entirety,
and the order in which data is received cannot be controlled.

Real time stream processing solutions are required
to manage this kind of data streams.
In fact, the generic platform for big data applications
proposed in~\cite{vilajosana2013}
assigns an important role to such a component.
Building scalable stream processing solutions
is far from trivial~\cite{cherniack03}.
In this paper we propose a system for scalably managing
streams of sensor data in the context of the
HERMES (\emph{Healthy and Efficient Routes
in Massive open-data basEd Smart cities})~\cite{hermes:jarca15} project,
which aims at helping its users, citizens of a smart city,
keep healthy habits.
Other systems for health care in smart cities
are reported in~\cite{solanas2014}.
The main sources of data in HERMES are the citizens themselves,
which contribute to the smart city by letting it track
their physical activities through activity bands
or the SmartCitizen mobile application,
and their driving through the SmartDriver mobile application.

In order to understand the amount of data it supposes,
let us focus on one of the applications.
The SmartDriver application aims at reducing
the stress levels and fuel consumption of its users,
as well as improving traffic safety,
by providing the user with real time driving
recommendations~\cite{Corcoba2015a,Corcoba2015b}.
In order to do that,
the application tracks its users while they drive
and sends the data to the infrastructure as soon as it captures it,
so that server-side services can perform real time computations
such as detection of congested roads and
stressful road sections.
The application should receive useful feedback back,
e.g. static road information,
a recommended driving speed and traffic alerts.
In its current prototype,
the application track's the vehicle's movement
as well as its driver's heart rate.
It reports the vehicle's location every 10 seconds.
In addition, it reports immediately abnormal situations
such as high accelerations or decelerations, excessive speeds
or abrupt increases in the driver's heart rate.
More detailed data, such as
second by second information about location, speed and heart rate,
are buffered in order to reduce resource consumption,
and sent to the infrastructure every time the driver
completes a 500 meter road section.
Because each driver produces at the least one data item every 10$s$,
only 10,000 drivers would suppose a load of more than
1,000 $requests/s$ for the infrastructure that collects the data.

The rest of the paper is organized as follows.
Section~\ref{sec:streaming-inf} proposes a system architecture
for the real time processing of streams in the context of a smart city.
Section~\ref{sec:case-study} presents a case study for this architecture,
based on the SmartDriver application of the HERMES project.
Section~\ref{sec:evaluation} reports the results of
a performance of the system.
Conclusions and future lines of work
are presented in section~\ref{conclusions}.

\section{The Data Streaming Infrastructure}
\label{sec:streaming-inf}

The server-side infrastructure was developed
on top of the Ztreamy middleware~\cite{fisteus2014}.
We have selected Ztreamy because of its flexibility,
scalability and the simple HTTP-based API it provides
to data producers,
which improves compatibility and simplifies the development of clients
such as the SmartDriver mobile application.
In addition, Ztreamy provides useful out-of-the-box features
such as stream aggregation, filtering and replication,
as well as a persistency subsystem that prevents the lose of
data items once they have been accepted by the infrastructure,
even in the case of temporal network disruptions
or failures of one or more components of the deployed system.
As our experiments in~\cite{fisteus2014} show,
other publish-subscribe systems for sensor data
like DataTurbine~\cite{fountain2009} would not provide
the performance levels we need in this scenario.
The ZeroMQ middleware\footnote{http://zeromq.org/ (Visited 2016-06-01)}
is more or less similar
in terms of performance to Ztreamy,
but Ztreamy provides us with a much more convenient high level API
and an HTTP-based interface.
The more recent Apache Kafka~\cite{kreps2011kafka} publish-subscribe system
could be an alternative to Ztreamy,
but we have not yet studied either its suitability for this scenario
or its performance,
and leave it for future work.

\begin{figure*}[t]
  \centering
  \includegraphics[width=0.65\textwidth]{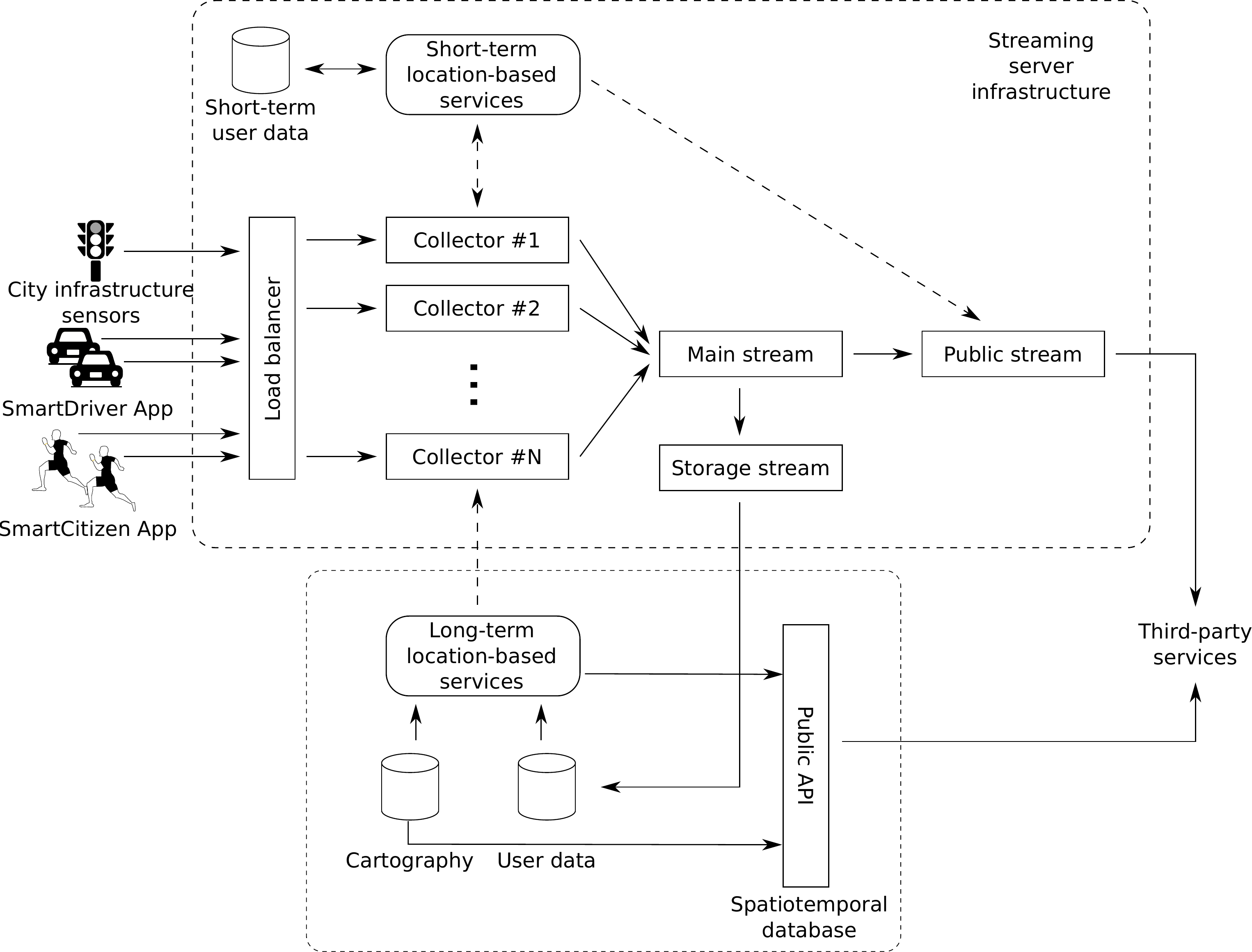}
  \caption{Data streaming infrastructure architecture}
  \label{fig:architecture}
\end{figure*}

Figure~\ref{fig:architecture} shows the system architecture
we have designed. It consists of the following main components:
\begin{itemize}
\item Data collectors:
  Ztreamy servers to which
  the SmartDriver and SmartCitizen mobile applications
  post their data through HTTP.
  These servers validate the data
  and orchestrate the interactions with other services
  needed to handle it.
  They are also responsible of responding mobile applications
  with feedback data when required.
  Since most of the load of input data handling
  is supported by these data collectors,
  they are replicated behind an HTTP load balancer
  in order to increase the number of clients they are able to handle.
  We have chosen the well-known
  Nginx\footnote{https://www.nginx.com/ (Visited 2016-11-23)}
  open-source HTTP server for this task.
\item Main stream:
  data items received by the collectors are then aggregated
  into the main stream, which is managed by a separate Ztreamy server.
\item Storage stream:
  this stream filters the data items that don't need to be stored
  out of the main stream.
  The HERMES servers that manage data persistence consume this stream
  in order to receive the data they have to store.
\item Public stream:
  this stream is derived from the main stream.
  It is part of the public API HERMES provides to third-party applications.
  It transports aggregated, anonymized and semantically-annotated data
  that may be useful
  to those applications.
\item Short-term location-based services:
  the streaming infrastructure needs to perform some real-time computations
  and keep some short-term data.
  For example, it needs to detect traffic incidents,
  retrieve the scores of nearby drivers
  for the SmartDriver's gamification system, etc.
  This module serves
  the collectors and the public stream server.
  In addition,
  it needs to use the long-term location-based services
  in order to get cartography data and
  speed recommendations based on historical data.
  This information is needed as input for some of the short-term services,
  and part of it is also returned to the SmartDriver application.

\end{itemize}

The other components of the architecture
(mobile applications, long-term storage and location-based services
and third-party applications) lay without the scope of this paper.

Depending on the amount of simultaneous clients the system needs to handle,
this architecture can be deployed on a single server
or distributed across several ones.
If distributed,
a good network link between them is advisable.
Ideally,
all the servers should share the same local network
in order to reduce end-to-end delays and bandwidth limitations.

Additionally,
because of the locality
of the services the infrastructure provides,
the system as a whole can be easily partitioned
for different geographical areas,
thus deploying a replica of the whole system for each geographical area.
This eases the scaling of the system
as the amount of users of its services grows.

\section{The SmartDriver Case Study}
\label{sec:case-study}

In order to illustrate the internals of the system,
let us focus on the SmartDriver mobile application.
It tracks the driver and posts the following types of events:
\begin{itemize}
\item \emph{Vehicle Location}:
  it contains a timestamp,
  latitude and longitude where the vehicle is located,
  an estimation of the accuracy of that location,
  the instantaneous vehicle speed
  and the current driving score assigned to the driver
  by the gamification subsystem of the application.
  These events are posted every 10$s$.
  They are used mainly for the real-time services.
\item \emph{Driving Section}:
  it contains more detailed information about a larger road section,
  including second by second location and speed,
  heart rate measurements
  and aggregated computations associated to this section
  (average and standard deviations of speed and heart rate,
  as well as statistics about speed variations).
  These events are posted for every 500$m$ the user drives.
  They are intended for storage,
  but can also be used in some real time services.
\item Abnormal situations:
  they are posted every time SmartDriver detects an abnormal situation
  (strong accelerations and decelerations,
  too high speeds,
  too high heart rates),
  immediately after its detection.
\end{itemize}

Because of their 10$s$ periodicity,
the system uses the \emph{Vehicle Location} posts
to send feedback to the SmartDriver application.
Collector servers are responsible of gathering the required information
from the short-term and long-term location-based services
and sending it back to the application in the body of their HTTP response.
This feedback includes:
\begin{itemize}
\item Type of road and its speed limit
  (to be obtained from the long-term services).
\item Recommended speed as computed by the speed recommendation service
  (to be obtained from the long-term services
  and possibly adapted to current road conditions by the
  short-term services).
\item Traffic alerts in the vicinity
  (to be obtained from the short-term services).
\item Driving scores assigned to nearby drivers by the gamification system
  (to be obtained from the short-term services).
\end{itemize}

In order to reduce the load of the long-term location-based services
with unnecessary requests due to stopped or very slow vehicles,
collectors assume the type of road and speed limit did not change
if the driver advanced less than 10$m$ since the last time
they determined those values.
The short-term services take similar measures to avoid
some computations such as retrieving or storing driver scores
in those situations.

The current prototype of the short-term location-based services
provides two main features:
\begin{itemize}
\item It tracks the latest location of each driver
  in order to detect the way of the road the driver follows
  (both the current location and a previous location
  are needed)
  as well as detecting when the driver has advanced
  more than the 10$m$ threshold.
\item It tracks the driving score and location of every driver
  in order to provide the nearby drivers' score service.
\end{itemize}

The first feature is implemented on top of a RAM-stored two-tier dictionary
in which every 30$s$ the oldest dictionary is dropped and a new one created.
This structure allows the system to keep just one location per driver
and drop those drivers that have not contacted the service
for more than 30$s$.

The second feature is more complex because it requires performing
spatial queries on a rectangle around the driver's current location.
We have implemented it with
a RAM-stored SQLite\footnote{https://www.sqlite.org/ (Visited 2016-06-01)}
database using an R-tree-based index.
The system periodically drops data older than 1 hour
because the gamification feature bases on recent data.

\section{Evaluation}
\label{sec:evaluation}

The current prototype
of the streaming server infrastructure
was subjected to experiments with varied amounts of load
in order to evaluate its performance.
Because of the unfeasibility of recruiting enough volunteers
to simultaneously use the application
up to the loads the system is able to handle,
we developed a simulator that produces a synthetic load.

\subsection{The Simulator}

The simulator was designed to produce data
and send it to the infrastructure
in a way that,
from the point of view of measuring performance,
is equivalent to having a given amount of actual users,
all of them using the SmartDriver application and driving simultaneously
a number of different paths in the same city.
The following parameters can be configured in the simulator
before starting a simulation:

\begin{itemize}
\item Number of simulated drivers:
  since each driver generates at least
  one \emph{Vehicle Location} event every 10s,
  the minimum number of requests per second the system needs to handle is
  $r_{min} = n / 10$, where $n$ is the number of drivers.
  \emph{Data Section} events make actual rates slightly higher,
  especially when drivers reach higher speeds.
  In order to introduce variability on the system,
  each driver is assigned some random parameters that influences
  her driver behavior
  (e.g. her inclination to drive fast or slow with respect to speed limits).
  In addition, not all drivers start at the same time.
  Each driver starts randomly within one minute of starting the simulator.

\item Paths:
  each simulated driver is assigned a path she will traverse
  during the simulation.
  Paths are based on the actual cartography of Seville,
  with random starting and end points in the city and its surroundings.
  Each path is created by choosing a pair of random start and end points
  within a configurable distance from the city center.
  The path itself will be the optimum path for going in a private vehicle
  from the start to the end point,
  as returned by a geographic information system.
  The number of paths is configurable
  and drivers are uniformly assigned to those paths.
  Therefore, many drivers may follow the same path.
  Despite sharing a path,
  because their behavior and the instant they begin to drive are random,
  those drivers will not be synchronized
  and therefore there will be enough randomness on the system.
\end{itemize}

Once the simulation starts,
the simulator makes every driver advance on her path
at a speed that depends on the randomly assigned characteristics of the driver
and the speed limit of the current road,
with a random bias.
Acceleration and deceleration are also modeled by the simulator
(e.g. at turns or when speed limits change).
Drivers send the events they produce to the infrastructure
by sending HTTP requests that are similar
to those the actual SmartDriver application would send.
Despite coming all the requests produced by the simulator
from the same host,
drivers in the simulator are prevented from sharing
their underlying TCP connections with other drivers.
This way the simulator will produce a realistic traffic pattern,
analogous to the actual pattern SmartDriver produces.

\subsection{Experimental Setup}

The experiments were run by deploying
the streaming server infrastructure
(load balancer, six collector instances,
one main stream instance,
one storage stream instance
and one short-term location-based service instance)
on a high-end server with 12 Intel Xeon E5-2430 2.5GHz cores
and 64 GB of RAM memory.

The simulator was deployed on a laptop computer,
connected to the server through one intermediate IP router
and a 100Mbps connection.
In order to accurately measure event delivery delays,
simulator and server used the NTP service to synchronize their clocks.

\subsection{Results}

\begin{figure}[t]
  \centering
  \resizebox{1.0\columnwidth}{!}{
      \includegraphics{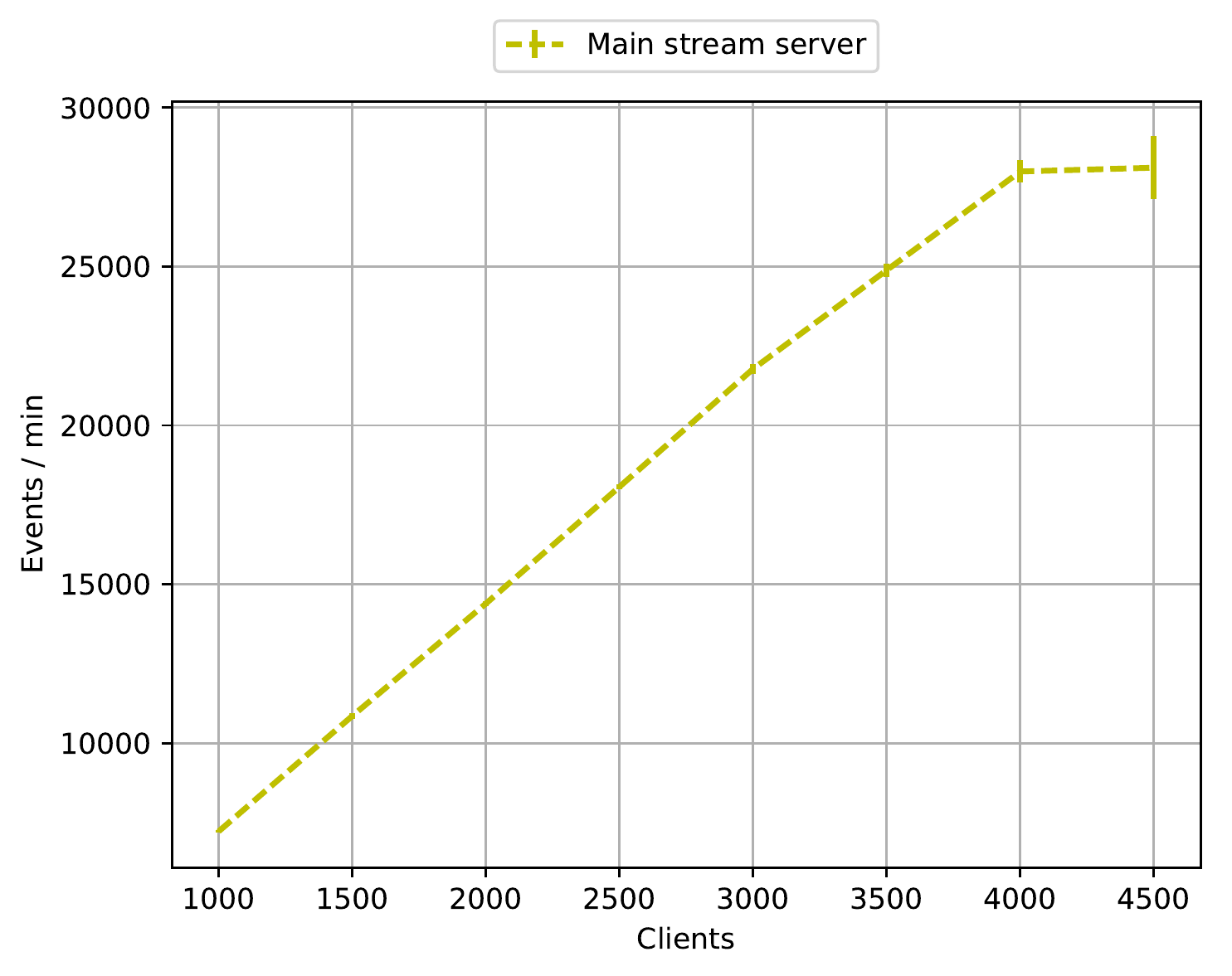}}
\caption{
  Evolution of event rate with respect to the number of clients.
  }
  \label{fig:event-rate}
\end{figure}

The combination of number of clients and event rate of each client
determines the load the system needs to handle.
Since in SmartDriver the event rate each client generates
is approximately the same,
the aggregated event rate arriving the server infrastructure
should grow linearly with the number of clients.
Figure~\ref{fig:event-rate} shows that, as expected,
the event rate at the main stream
is proportional to the number of clients
up to 4,000 clients.
At that point,
with approximately 28,000 events per minute,
the infrastructure saturates and begins to reject events,
and therefore linearity is lost.

We've measured the performance of the server infrastructure
for different loads
in a series of experiments with a growing number of clients.
The main performance indicators we measured were:
\begin{itemize}
\item CPU utilization:
  amount of time of CPU used
  during a minute, divided by $60s$.
  This measurement was taken every minute.
  Its estimated mean and 95\% confidence intervals
  were computed and reported in the plots.
  A component with an utilization close to $1$
  is in the limit of the load it can handle.
  A component with an utilization close to $0$ is mainly free.
\item Event admission delay:
  amount of time between the creation of the event at the client side
  (the simulator)
  and its admission at a front-end server
  and at the database feed stream.
  Larger delays may signal congestion situations in the server.
  Similarly to utilization,
  mean delays with 95\% confidence intervals were estimated
  from the delays suffered by a random sample
  of the simulated events.
\end{itemize}

\begin{figure}[t]
  \centering
  \resizebox{1.0\columnwidth}{!}{
      \includegraphics{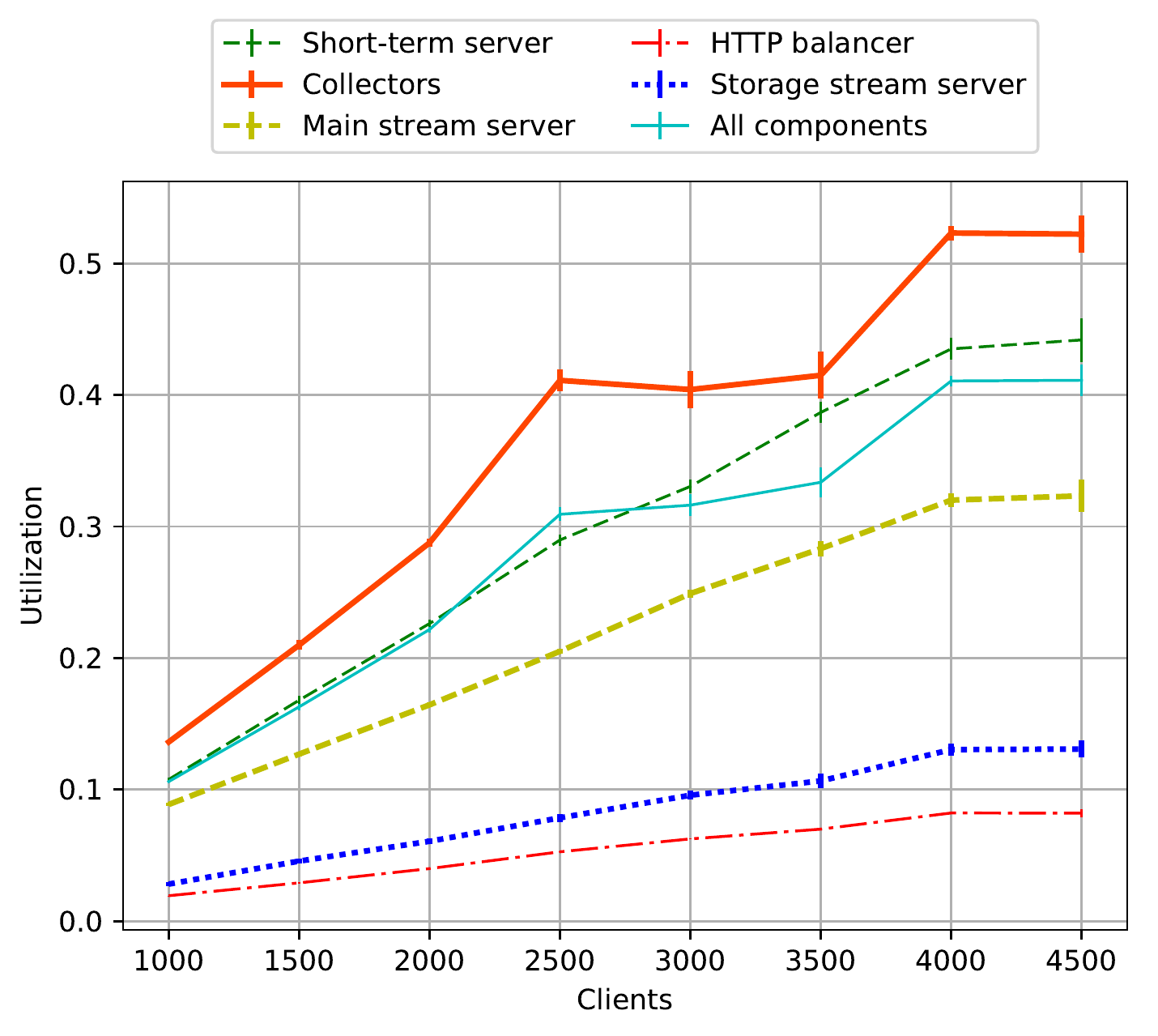}}
\caption{
  Global and per-component CPU utilization.
  }
  \label{fig:utilization}
\end{figure}

\begin{figure}[t]
  \centering
  \resizebox{1.0\columnwidth}{!}{
      \includegraphics{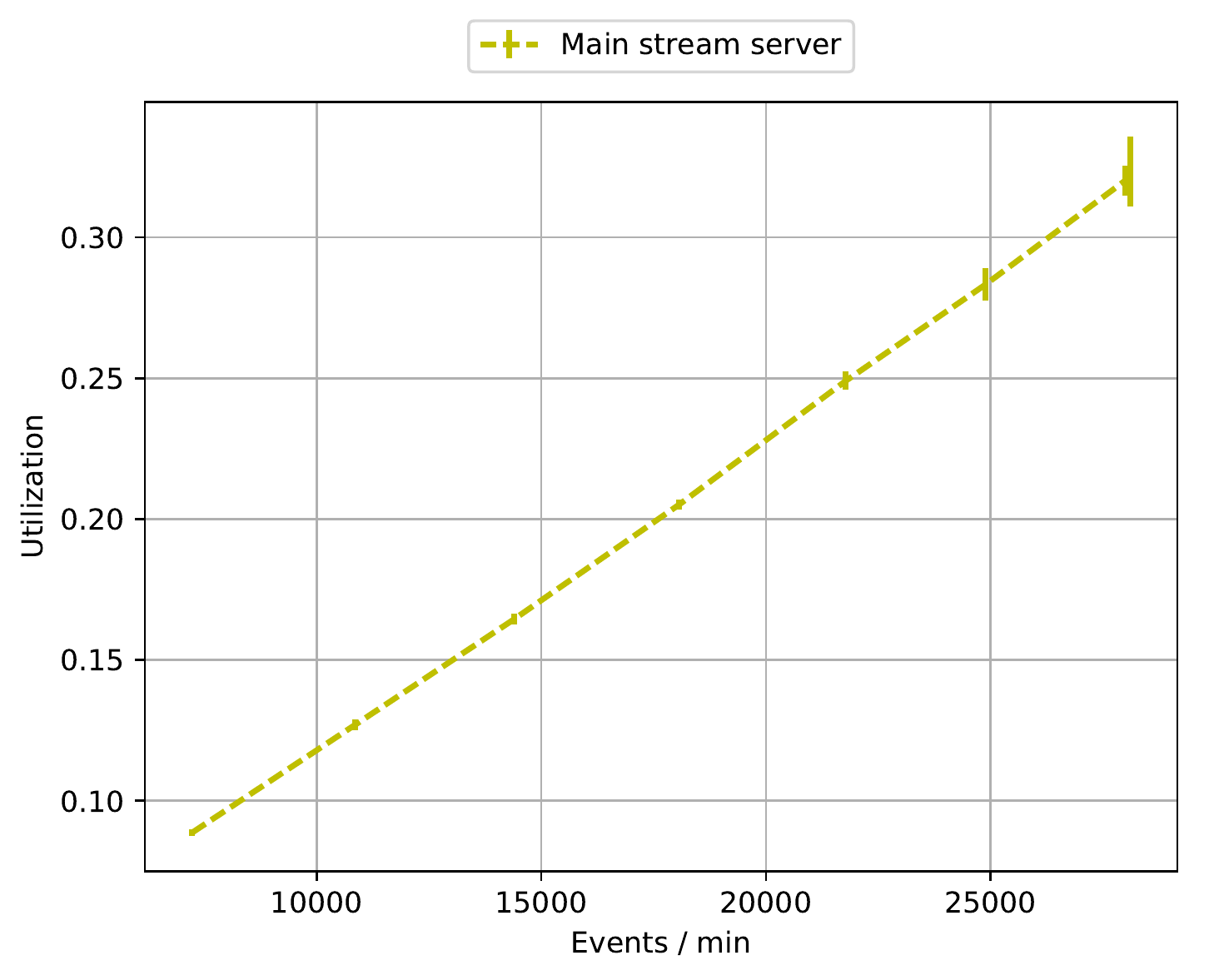}}
\caption{
  Utilization versus event rate.
  }
  \label{fig:utilization-events}
\end{figure}

Figure~\ref{fig:utilization} shows the overall utilization
of the infrastructure as well as the individual utilization
of each server component.
The six collector servers average a higher utilization
than the rest,
thus being the bottleneck of the system.
However, their utilization may be reduced
by adding some more collector instances to the pool,
assuming that the server has cores enough.
The next component in terms of utilization
is the short-term location-based server,
followed by the server that handles the main stream.
The storage stream server and the HTTP balancer
can handle much more load than the rest.
Figure~\ref{fig:utilization-events} shows clearly
that utilization at the main stream grows linearly with its event rate.

\begin{figure}[t]
  \centering
  \resizebox{1.0\columnwidth}{!}{
      \includegraphics{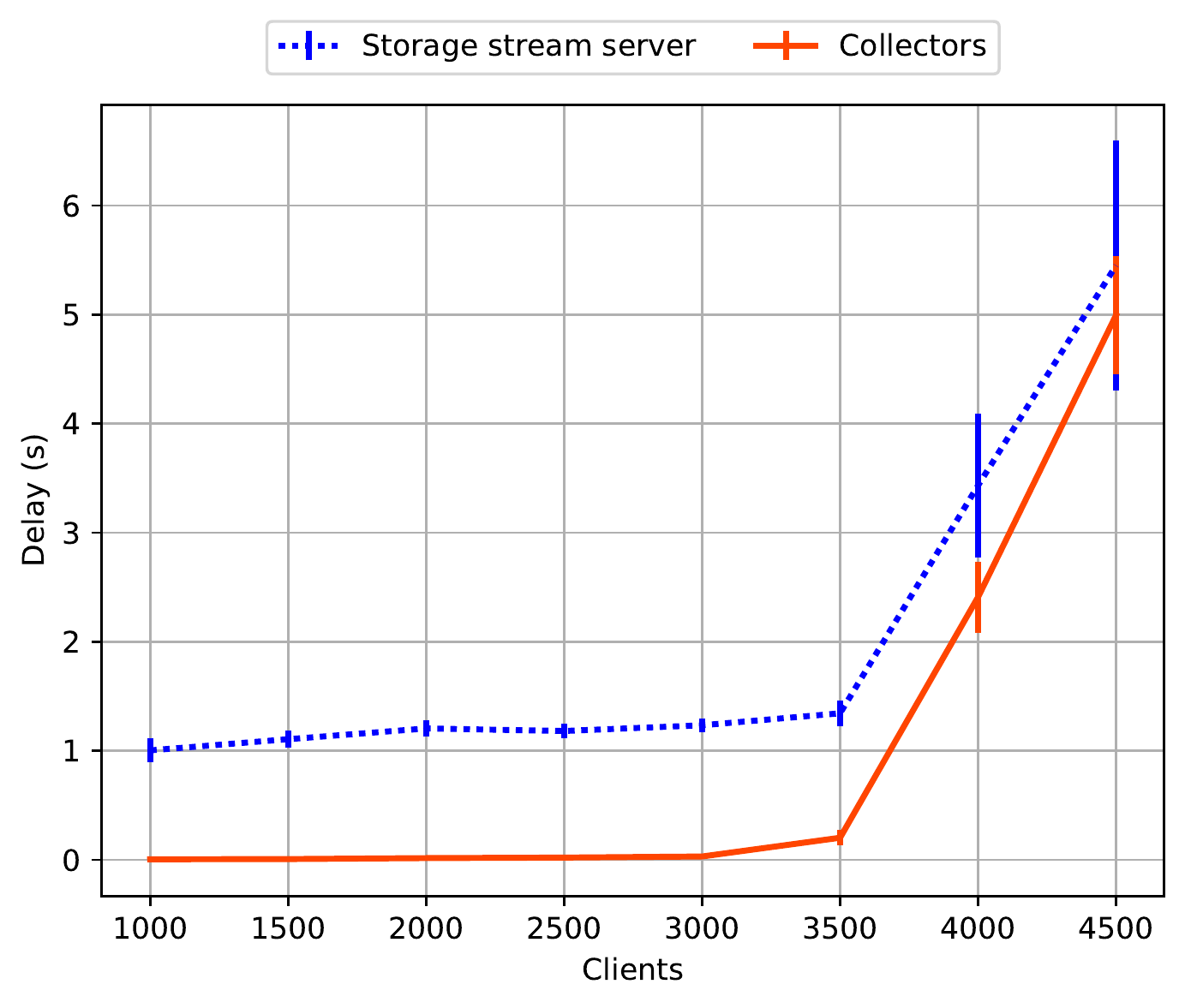}}
\caption{
  Event delivery delays at collectors and storage stream.
  Delays measure the amount of time since an event is created
  at the simulator
  until it is fully processed and accepted at the given server component.
  }
  \label{fig:delays}
\end{figure}

Finally, figure~\ref{fig:delays} shows the event delivery delay at two points
of the server infrastructure:
collectors and storage feed stream.
As expected, delays increase with the load of the system,
especially from 3,500 clients on,
where the indicators show that the system is beginning to saturate.
In our experimental setup the network distance
between the simulator and the infrastructure
is less than $1ms$.
In more realistic scenarios,
that distance would be higher,
although less than $0.5s$ in most of the situations.

As a conclusion,
the infrastructure we have presented in this paper
is able to handle up to 4,000 simultaneous drivers from a single server,
which represent approximately 28,000 new events every minute.
At larger data rates collectors begin to reject some events due to saturation.

\section{Conclusions and Future Work}
\label{conclusions}

We deployed the first working prototype of this infrastructure
more than two years ago.
It is still working
and has received frequent feature upgrades and bug fixes since then.
During this period,
the system has been continuously capturing data
from our beta testers with no major issues.
Although the core of the architecture is already implemented and deployed,
some of its services are still work in progress.
More specifically,
the public stream and
the speed recommendation and traffic incident detection services
are not yet part of the current prototype.

According to our experiments,
the architecture we propose
provides a reasonable level of performance
in the context of the HERMES project.
The maximum amount of simultaneous drivers a single server may handle
is approximately 4,000,
but the infrastructure can be scaled-up by deploying it into more servers,
especially the collectors components.
Distributing the short-term location services component
is challenging because of the need of a shared spatial database,
but techniques for efficiently partitioning such data
and their processing already exist~\cite{aji2013,wang2015}.

Future work includes rebuilding this infrastructure
on top of a big data framework
such as Apache Kafka~\cite{kreps2011kafka}
to try to increase the amount of clients that may be served
from a single server.

\subsubsection*{Acknowledgements}

Research reported in this paper was supported by the Spanish Economy
Ministry through the ``HERMES -- Smart Driver'' project
(TIN2013-46801-C4-2-R)
and the ``HERMES -- Smart Citizen'' project
(TIN2013-46801-C4-1-R).

%% The file named.bst is a bibliography style file for BibTeX 0.99c
%\bibliographystyle{named}

% \bibliographystyle{alpha}
% \bibliography{ceur}

\newcommand{\etalchar}[1]{$^{#1}$}

\end{document}